\documentclass[twocolumn,showpacs,preprintnumbers,amsmath,amssymb]{revtex4}
\usepackage{graphicx}
\usepackage{dcolumn}
\usepackage{bm}

\begin{document}

\title{In-beam fast-timing measurements in $^{103,105,107}$Cd} 

\author{S.~Kisyov$^1$, S.~Lalkovski$^{1}$\footnote{E-mail address: stl@phys.uni-sofia.bg},
N.~M\v{a}rginean$^2$, D.~Bucurescu$^2$, L.~Atanasova$^3$, 
D.~L.~Balabanski$^3$, Gh.~C\v{a}ta-Danil$^2$, I.~C\v{a}ta-Danil$^2$, 
J.-M.~Daugas$^4$, D.~Deleanu$^2$, P.~Detistov$^3$, D.~Filipescu$^2$, 
G.~Georgiev$^5$, D.~Ghi\c{t}\v{a}$^2$, T.~Glodariu$^2$, J.~Jolie$^6$, 
D.S.~Judson$^7$, R.~Lozeva$^5$, R.~M\v{a}rginean$^2$, C.~Mihai$^2$, 
A.~Negret$^2$, S.~Pascu$^2$, 
D.~Radulov$^1$\footnote{Present address: Katholieke Universiteit, Leuven, Belgium}, J.-M.~R\'egis$^6$, M.~Rudigier$^6$, 
T.~Sava$^2$, L.~Stroe$^2$, G.~Suliman$^2$, N.V.~Zamfir$^2$, K.O.~Zell$^{6}$ 
and M.~Zhekova$^{1}$
}

\affiliation{
$^1$Faculty of Physics, University of Sofia "St.~Kliment Ohridski", 1164 Sofia, Bulgaria; \\
$^2$Horia Hulubei National Institute for Physics and Nuclear Engineering, 
77125 Bucharest-Magurele,
Romania \\
$^3$Institute for Nuclear Research and Nuclear Energy, Bulgarian
Academy of Science, 1784 Sofia, Bulgaria \\
$^4$CEA, DAM, DIF, 91297 Arpajon, France\\
$^{5}$Centre de Spectrom\'etrie Nucleaire et Spectrometrie de Masse,
91405 Orsay-Campus, France\\
$^6$ Institut f\"ur Kernphysik, University of Cologne, Cologne, Germany\\
$^7$Department of Physics, University of Liverpool, Liverpool,
United Kingdom\\
}

\begin{abstract}
Fast-timing measurements were performed recently in the region of the 
medium-mass $^{103,105,107}$Cd isotopes, produced in fusion evaporation 
reactions. Emitted $\gamma$-rays were detected by eight HPGe and 
five LaBr$_3$:Ce detectors working in 
coincidence. Results on new and re-evaluated half-lives are discussed within
a systematic of transition rates. The $7/2_1^+$ states in $^{103,105,107}$Cd 
are interpreted as arising from a single-particle excitation. The
half-life analysis of the $11/2_1^-$ states in  $^{103,105,107}$Cd
shows no change in the single-particle transition strength 
as a function of the neutron number.   
\end{abstract}

\date{\today}
\maketitle

\section{Introduction}
Cadmium isotopes have two protons 
less than the $_{50}$Sn nuclei, presenting a good test case for the
robustness of the shell structure.
Shell model calculations successfully describe the experimentally observed 
level energies and level lifetimes in the extreme neutron-rich and 
neutron-deficient cadmium isotopes proving the persistence of the shell 
structure below the doubly magic $^{132}$Sn and $^{100}$Sn  
\cite{Ju07, Ca09, Na10, Bl04}. Fingerprints of collectivity, 
however, start to emerge when moving away from the neutron shell closures. 
They can be found in the decrease of the $2_1^+$ energy and in the increase of the 
respective B(E2;$2_1^+\rightarrow 0_1^+$) values when approaching the neutron mid-shell
\cite{nndc}. 

Due to the neighbourhood of the shell model tin isotopes
and the presence of weak collectivity in the neutron-mid shell cadmium isotopes, both 
single particle and collective states are expected to occur in the medium mass 
odd-A Cd nuclei. Moreover, there are several cases where the structure of the 
state is ambiguous. In the $^{103,105,107}$Cd \cite {Fr09, Fr05, Bl08}, 
for example, the lowest-lying excited $J^\pi=7/2^+$ state can arise from a collective 
excitation built on the $5/2^+$ ground state or from a single-particle excitation. 
A model independent approach to the problem is to evaluate the $B(E2)$ transition 
strengths within a systematical study involving even-even well deformed and spherical 
nuclei, where the structure is well established.

In order to study the structure of the low-lying excited states in $^{103,105,107}$Cd 
fast-timing measurements were performed. The half-lives are 
directly related to the transitions rates and hence to the structure of the state. 
The present paper reports on new results, obtained with eight HPGe 
detectors working in coincidence with five LaBr$_3$:Ce detectors. 

\section{Experimental Set Up}
The low-lying excited states, placed on and close to the yrast line 
in $^{103}$Cd, $^{105}$Cd and $^{107}$Cd were populated via fusion 
evaporation reactions. A carbon beam, accelerated to 50 MeV by the 
Tandem accelerator of the National Institute for Physics and 
Nuclear Engineering at Magurele, Romania, impinged on
self-supporting 10 mg/cm$^{2}$ thick $^{94,96}$Mo targets and on a 
1 mg/cm$^{2}$ thick $^{98}$Mo target with 20 $\mu$m Pb backing. The three 
targets were isotopically enriched up to 98.97\% in $^{94}$Mo, 
95.70\% in $^{96}$Mo and 98\% in $^{98}$Mo, respectively. 

The cross section for the $^{94}$Mo($^{12}$C,3n)$^{103}$Cd reaction was 
calculated to be 100 mb, while for the 
$^{96}$Mo($^{12}$C,3n)$^{105}$Cd and $^{98}$Mo($^{12}$C,3n)$^{107}$Cd 
reactions it was approximately 400 mb. The typical beam intensity was 
of the order of 8 pnA. Besides the 3n channels, the 4n, 2np, 4np and 
2n$\alpha$- fusion evaporation channels also have significant 
cross sections which contaminate the spectra of interest.

The half-lives of the levels of interest were deduced by using a fast-timing 
set up consisting of 5 LaBr$_3$:Ce scintillator detectors working in 
coincidences with 8 HPGe detectors \cite{Ma10}. Five of the HPGe detectors 
were placed at backward angles with respect to the beam axis, two 
were placed at 90$^\circ$
and the eighth HPGe detector was placed at a forward angle. The five
LaBr$_3$:Ce detectors were mounted bellow the target chamber on a ring of
approximately 45$^\circ$ degrees with respect to the beam axis.  
The five LaBr$_3$:Ce crystals had a cylindrical shape and 
5\% Ce doping. One of the LaBr$_3$:Ce detector was a commercial 
integral detector. Its size was 2''$\times$2''. Two of the LaBr$_3$:Ce 
detectors had 1'' height and a diameter of 1''. Two LaBr$_3$:Ce crystals 
had dimensions of 1.5''$\times$1.5''. Each of the four crystals was optically 
coupled to XP20D0B photomultiplier and mounted in aluminum casing. 
The readout, from each of the four non-comercial detectors, was made 
via a VD184/T  voltage divider. The voltage divider issues 
a negative anode signal and a fast positive dynode signal. 
The anode signal was used for timing, while the dynode signal was used to 
obtain energy signal. This non conventional choice was made to 
avoid the saturation of the dynode signal \cite{Ma10}, which facilitates 
the analysis of the energy spectra.

The energy signals from the HPGe detectors were amplified and then 
digitized by 8k Analog to Digital Converters 
(ADC) AD413A. The timing signals from the HPGe detectors 
were processed by 4k 4418/T Time-to-Digital converters. 
The energy signals from the LaBr$_3$:Ce detectors were 
amplified by spectroscopic amplifiers and then 
digitized by 8k ADC AD413A. The timing signals from the LaBr$_3$:Ce 
detectors were sent to a Quad Constant Fraction Discriminator, model 935. 
Each of the five timing signals was used to start a Time-to-Amplitude 
Converter (TAC) operating in a common stop mode. Then the five TAC output 
signals were sent to 8k ADCs. The acquisition was triggered 
when two LaBr$_3$:Ce and one HPGe detectors were fired in coincidences. 

\section{Data Analysis}
Data was stored in event-by-event mode in 100 MB long files, which were 
grouped in runs of approximately 2 hours. Then the data was 
analyzed using the GASPWare and Radware \cite{Ra95} packages.  
Because of the instability of the LaBr$_3$:Ce detectors observed with 
time, a gain matching procedure was applied run-by-run.
To correct the CFD for the walk effect, observed at low energies, 
analysis of the time responce as a function of energy was 
performed with a $^{60}$Co source \cite{Ma10} and in-beam. Then the data 
was sorted in gated energy spectra, two-dimentional 
energy-energy ($E_\gamma-E_\gamma$) and three-dimentional energy-energy-time 
($E_\gamma-E_\gamma-\Delta T$) matrices, where $E_\gamma$ is the $\gamma$-ray
energy detected by a LaBr$_3$:Ce detector and $\Delta T$ is a
time difference between two gamma rays detected in coincidence.

The ($E_\gamma-E_\gamma-\Delta T$) matrices were constructed
as fully symetric in energy, i.e. for each event where $\gamma$-rays of 
energies $E_{\gamma 1}$ and $E_{\gamma 2}$ are detected the matrix elements 
($E_{\gamma 1}, E_{\gamma 2}$) and ($E_{\gamma 2}, E_{\gamma 1}$) are incremented, 
while the time intervals associated with these two points are calculated as 
$\Delta T= (t_1-t_2)+t_0$ and $\Delta T= -(t_1-t_2)+t_0$ respectively. 
Here, $t_1-t_2>0$ is the time 
difference measured with two TAC converters and $t_0$ is an arbitrary offset.
In the cases where the two $\gamma$-rays feed and de-excite a state with a 
half-life longer than the electronics resolution, which in the present work 
is 6 ps/channel, then the time distributions associated with the two matrix elements 
($E_{\gamma 1},E_{\gamma 2}$) and ($E_{\gamma 2},E_{\gamma 1}$) will be shifted 
by $2\tau$, where $\tau$ is the lifetime of the level of interest. 
This procedure represents the centroid shift 
method \cite{An82}, which has been successfully used in the past \cite{An85} 
and recently applied with LaBr$_3$:Ce detectors \cite{Ma10}. In the cases, 
where the level half-life is much longer than the detector time resolution,
a tail emerges on the right hand side of the time distribution. In these cases 
the slope of the tail has been used to determine the half-life of the level.
Deconvolution of Gaussian and exponent was applied in the cases where the half-life
of the level is of the order of the FWHM of the prompt distribution.

In order to select a particular reaction channel and particular $\gamma$-decay 
branch leading to the state of interest, the  matrices were constructed 
with a condition imposed on prompt $\gamma$-rays detected in any of the 
high-resolution HPGe detectors. 

\begin{figure}
\rotatebox{-90}{\scalebox{0.35}[0.35]{\includegraphics{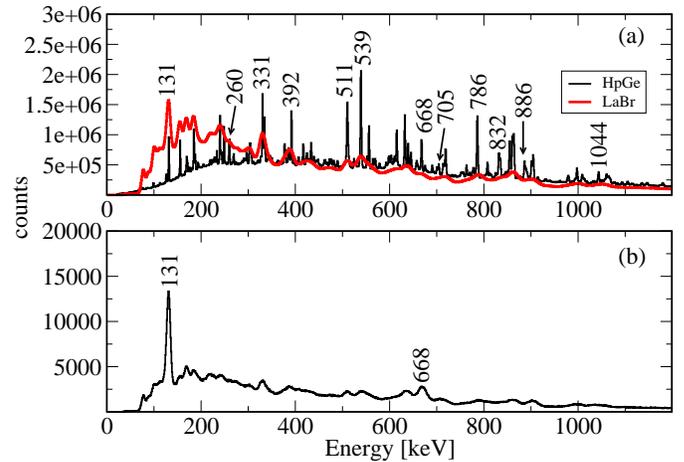}}}
\caption{(Colour online) $^{105}$Cd energy spectra: (a) Total projection from all HPGe and LaBr$_3$:Ce detectors; (b) LaBr$_3$:Ce energy spectrum gated on a 886-keV transition in any of the HPGe detectors. The labels denote transitions in $^{105}$Cd}
\label{ene}
\end{figure}

Fig.~\ref{ene}(a) shows the energy total projection for the  
$^{12}$C+$^{96}$Mo$\rightarrow ^{105}$Cd+3n reaction for all 
HPGe and LaBr$_3$:Ce detectors. At low energies, the higher 
efficiency of the LaBr$_3$:Ce with respect to the HPGe detectors
is remarkable. The energies of $^{105}$Cd are marked with numbers. 
Fig.~\ref{ene}(b) represents the LaBr$_3$:Ce energy spectrum, gated on 
the 886-keV transition from $^{105}$Cd in the HPGe detectors, 
which improves the peak-to-background 
ratio. Similar spectra were constructed for the other two reactions
$^{12}$C+$^{94}$Mo$\rightarrow ^{103}$Cd+3n and 
$^{12}$C+$^{98}$Mo$\rightarrow ^{107}$Cd+3n. 

\begin{figure}
\rotatebox{-90}{\scalebox{0.35}[0.35]{\includegraphics{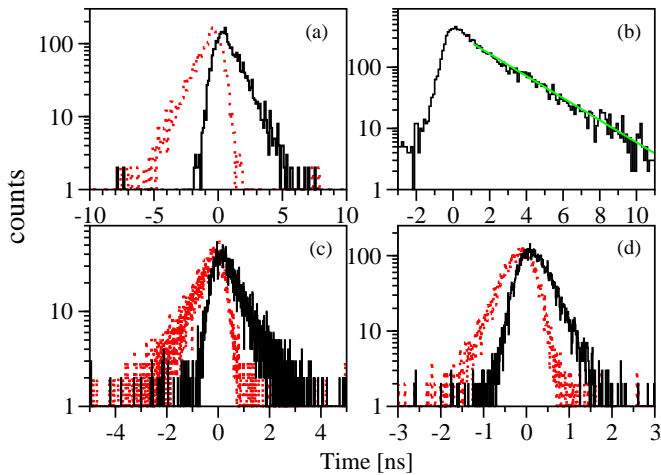}}}
\caption{(Colour online) Time spectra, obtained for the decay of the $7/2_1^+$ states in $^{107}$Cd (a), $^{105}$Cd (b), $^{103}$Cd (c) and for the $11/2_1^-$ state in $^{105}$Cd (d)}
\label{times}
\end{figure}

\begin{figure}
\rotatebox{90}{\scalebox{0.30}[0.30]{\includegraphics{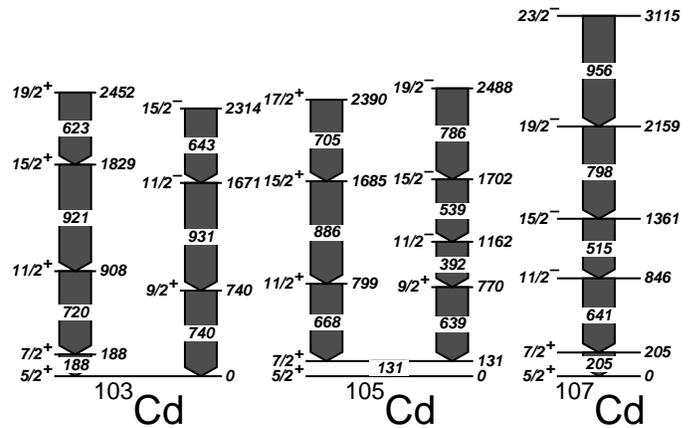}}}
\caption{Partial level schemes of $^{103,105,107}$Cd}
\label{part}
\end{figure}

Fig.~\ref{times} presents time spectra obtained after two dimensional energy 
gates imposed on $E_\gamma$-$E_\gamma-\Delta T$ matrices, gated on prompt transitions
with HPGe detectors. To increase the statistics, in each of the cases, 
several prompt gates were imposed on the HPGe detectors. Here, the procedure
will be ilustrated by using the lowest lying prompt and delayed transitions shown
on Fig.~\ref{part}.

Fig.~\ref{times}(a) presents the time distributions for the decay 
of the $7/2_1^+$ state in $^{107}$Cd. The time distribution, plotted in full 
lines, is obtained with a ($205\gamma, 641\gamma$) energy gate, while the 
symmetric ($641\gamma, 205\gamma$) gate is plotted with dots. The half-life 
of 0.68 (4) ns, obtained from the centroid shift method, is consistent with the 
NNDC value of $T_{1/2}=$0.71 (4) ns \cite{Bl08}. Gates on 798-keV or 956-keV 
transitions (Fig.~\ref{part}) were  applied with HPGe detectors in order to 
clean the time spectra from background events. 

Fig.~\ref{times}(b) presents the time curves for the decay of the $7/2_1^+$ 
state in $^{105}$Cd. The half-life of 1.66 (12) ns, obtained in the present 
study, was measured from the slope of the time distribution gated on 
(639$\gamma$-131$\gamma$) with the LaBr$_3$:Ce 
detectors and cleaned with a gate on the 886 $\gamma$-ray or 705-keV $\gamma$-ray 
(Fig.~\ref{part}) imposed on any of the eight HPGe detectors. 
It agrees the 1.75 (11) ns value, adopted by NNDC \cite{Fr05}, which 
is based on a $\gamma (t)$ measurement with one NaI(Tl) detector \cite{Ro73}. 

Fig.~\ref{times}(c) presents the time curves for the decay of the $7/2^+_1$ 
state in $^{103}$Cd. The half-life of 0.37 (3) ns was obtained from the 
centroid shift of the two time distributions generated with gates on the 
188-keV and 720-keV transitions (Fig.~\ref{part}) imposed on any two of the 
LaBr$_3$:Ce detectors in coincidence with the 921-keV or 623-keV $\gamma$-rays 
(Fig.~\ref{part}) detected in any of the HPGe detectors.

The half-life of the $11/2_1^-$ state in $^{105}$Cd Fig.~\ref{times}(d)
was obtained by gating on the 539-keV feeding and 392-keV de-exciting 
transitions (Fig.~\ref{part}), detected by any two of the five LaBr$_3$:Ce 
detectors. An additional gate on the 786-keV $\gamma$-ray, which is in 
coincidence with the 392-keV and 786-keV transitions (Fig.~\ref{part}), 
was imposed on any of the HPGe detectors. The half-life, deduced from 
the centroid shift of the two mirror time spectra, is 149 (12) ps.

\section{Discussion}
\begin{table}
\caption{
First excited $7/2^+$ state in $^{103,105,107}$Cd and decay properties}
\label{threecad}
\begin{ruledtabular}
\begin{tabular}{ccccccccccc}
Isotope & $E_i$ & $T_{1/2}$ & $E_\gamma$ 
& $L\lambda$ & $\delta$ &  $B(\lambda L)$\\
        & [keV]              &       & [keV] &  & & [W.u.]\\
\hline
$^{103}$Cd & 188  &  0.37 (3) ns & 188  & M1      & $\leq 0.1$ & 0.0089 (8) \\
           &     &           &      & E2      &            & 2.27 (19) \\  
\hline
$^{105}$Cd  & 131  & 1.66 (12) ns & 131 &  M1   & $\leq 0.1$    & 0.0058 (5)\\ 
            &       &            &     & E2   &              &  2.93 (22)\\ 
\hline
$^{107}$Cd  & 205 &   0.68 (4) ns  & 205  & M1     & 0.25 (1) & 0.00331 (20)\\
            &            &             &      & E2     &         & 4.2 (4)\\
\end{tabular}
\end{ruledtabular}
\end{table}

$\mathbf {7/2_1^+:}$ The half-lives $T_{1/2}$, of the levels of interest, 
are listed in Table~\ref{threecad} along with the level energy $E_i$ and 
the spin/parity assignments $J^\pi$. In order to calculate the partial 
half-lives and the reduced transition probabilities the $\gamma$-ray 
energies $E_\gamma$, multipolarities $L\lambda$, and mixing ratios $\delta$, 
adopted by NNDC \cite{Fr09, Fr05, Bl08}, are also listed. The $J^\pi=7/2^+$ 
state is the first excited state in all three isotopes and decays via 
M1+E2 transition to the ground state. An upper limit of the mixing ratio 
$\delta \leq 0.1$ for the $7/2_1^+\rightarrow 5/2_1^+$ transition
in $^{103}$Cd has been estimated by the NNDC \cite{Fr09}. 
The mixing ratio, adopted for the M1+E2 transition in $^{107}$Cd, is 
$\delta =$+0.25 (1) \cite{Bl08}. It has been 
suggested that the respective transition in $^{105}$Cd is of almost pure M1 
nature, however a small E2 admixture is assumed \cite{Fr05}. For the purpose 
of the current discussion an upper limit of $\delta \leq$0.1 was adopted 
in the present study.

\begin{figure}[t]
\rotatebox{-90}{\scalebox{0.35}[0.35]{\includegraphics{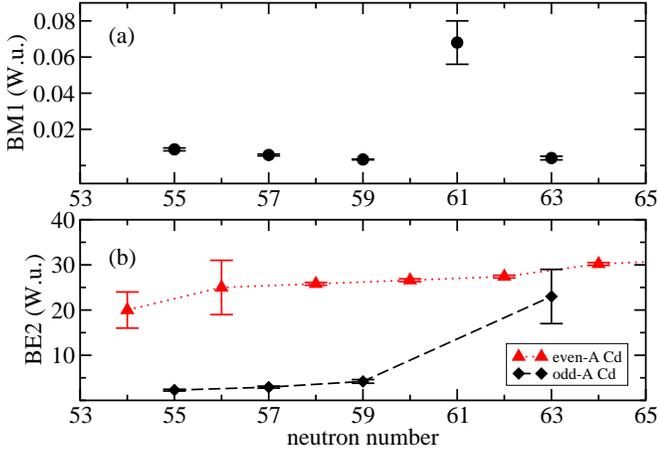}}}
\caption{(Colour online) Systematics of (a) B(M1;$7/2_1^+\rightarrow 5/2_1^+$) values for odd-A cadmium nuclei and (b) B(E2;$7/2_1^+\rightarrow 5/2_1^+$)  values for odd-A cadmium isotopes (diamonds) compared to the B(E2;$2_1^+\rightarrow 0_1^+$) values in even-even cadmium cores (triangles)}
\label{BM1BE2}
\end{figure}

The reduced transition probabilities, calculated with RULER \cite{nndc}, 
are given in the last column of Table~\ref{threecad}.

\begin{figure}[t]
\rotatebox{-90}{\scalebox{0.35}[0.35]{\includegraphics{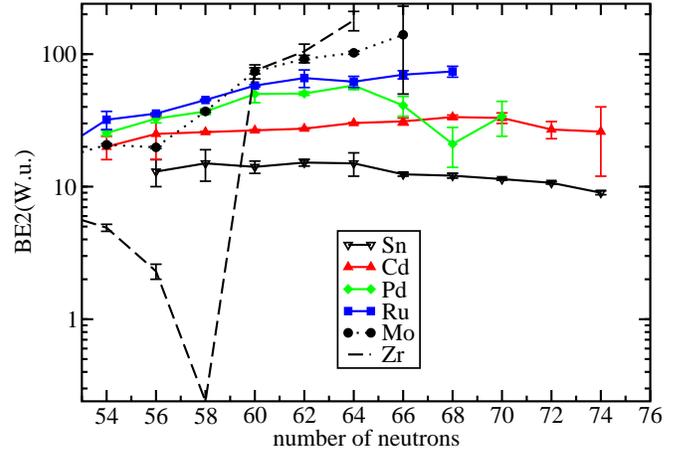}}}
\caption{(Colour online) Systematics of B(E2;$2_1^+\rightarrow 0_1^+$) transition rates for the even-even nuclei with 40$\leq$ Z $\leq 50$}
\label{BE2}
\end{figure}

Fig.~\ref{BM1BE2} shows the systematic trend of the B(M1) 
(Fig.~\ref{BM1BE2}(a)) and B(E2) values for the $7/2_1^+\rightarrow 5/2^+_1$ 
transitions in $^{103-111}$Cd$_{55-63}$ (Fig.~\ref{BM1BE2}(b)), compared to the 
B(E2;$2_1^+\rightarrow 0_1^+$) for their even-even Cd cores. Fig.~\ref{BE2} 
shows the evolution of the B(E2;$2_1^+\rightarrow 0_1^+$) transition rates 
with the neutron number for all even-even nuclei in the $40\leq$Z$\leq 50$ 
region.

In the $^{103-107}$Cd$_{55-59}$ nuclei, because of the low mixing ratio, 
the B(E2;$7/2_1^+\rightarrow 5/2^+_1$) transition strenghts are 
significantly suppressed in comparison to the B(E2;$2_1^+\rightarrow 0_1^+$) values
for the even-even Cd cores (Fig.~\ref{BM1BE2}(b)). Moreover, they are two orders 
of magnitude weaker than the B(E2;$2^+_1\rightarrow 0^+_1$) values for the 
most deformed neutron mid-shell Zr and Mo nuclei (Fig.~\ref{BE2}). In fact, 
the $^{103,105,107}$Cd
B(E2;$7/2_1^+\rightarrow 5/2_1^+$) values are similar to the reduced 
transition probabilities for the magic tin nuclei (Fig.~\ref{BE2}) suggesting a 
single-particle nature of the $7/2_1^+$ state, most probably arising 
from $\nu g_{7/2}$ configuration.
 
In $^{109}$Cd$_{61}$, the $7/2_1^+$ state appears 203 keV above the $5/2^+$ 
ground state and decays via a pure, according to NNDC, M1 transition 
giving rise to B(M1)(W.u.)=$0.068$ 12 \cite{Bl06}, which is an order of magnitude
higher than the respective value in $^{103-107}$Cd. However, the odd behavior 
of the B(M1) point on Fig.~\ref{BM1BE2} suggests a significant E2 component.
In fact, such an increase of the $B(E2)$ value, and hence in the collectivity of 
the state, is observed in $^{111}$Cd. There, the $5/2_1^+$ and $7/2_1^+$ states 
appear at 245 keV and 416 keV respectively \cite{Bl09}. The $7/2_1^+$ has a 
half-life of 0.12 ns and decays to the $5/2_1^+$ state via 171-keV M1+E2 transition.
The mixing ratio $\delta = -0.144$ of this transition leads to 
B(E2;$7/2_1^+\rightarrow 5/2_1^+$)=23 6 (2) W.u., which approaches the  
B(E2;$2_1^+\rightarrow 0_1^+$) value in the even-even cadmium cores 
(Fig.~\ref{BM1BE2}(b)) and hence the $7/2^+$ state becomes collective.

$\mathbf {11/2_1^-:}$ The $J^\pi = 11/2_1^-$ states appear in all 
odd-cadmium isotopes
from $^{103}$Cd to $^{123}$Cd \cite{nndc}. It is observed at 1671 keV in 
$^{103}$Cd \cite{Fr09} and decreases in energy when approaching the neutron 
mid-shell. In $^{117}$Cd$_{69}$ it appears at 136 keV above the 
ground state. 

The big energy gap between the $11/2_1^-$ state and the 
ground state in $^{103-111}$Cd opens space for 
several states of single-particle and collective nature
to appear. Among those levels is the $9/2_1^+$ state to which
the $11/2_1^-$ state decays via an E1 transition. For medium mass 
odd-A Cd isotopes the energy of the $11/2_1^-$ drops closer to the 
ground state where only low-spin states are populated. In this mass 
region the $11/2_1^-$ state decays via low-energy transitions of higher
multipolarity leading to increase of its half-life to $T_{1/2}=$14.1 y
in $^{113}$Cd \cite{nndc}. Further increases of the energy of the 
$11/2_1^-$ state 
with respect to the ground state leads to a decrease of the half-life. 
The isomeric $11/2_1^-$ state in all neutron-rich cadmium nuclei with 
A$\geq$113 decay  
via $\beta$-decay process to the respective indium isobars. In spite
the half-life of the state decreases and the energy of the isomeric 
state increases with the mass number, isomeric decays have not been 
observed so far.

The half-life of the excited $11/2_1^-$ state in $^{103,105}$Cd, measured
in present work, allows a systematical study of the E1 transitions 
strenghts as a function of the neutron number. The half-life $T_{1/2}$=71 ns 
of the $11/2_1^-$ state in $^{107}$Cd has been previously measured \cite{Ro73}.
This level decays via a branch of E1, M2 and E3 transitions to $9/2^+$,
$7/2^+$ and $5/2^+$ states with partial half-lives 2.4$\times 10^{-7}$,
1.0$\times 10^{-7}$ and 4.5$\times 10^{-6}$ s respectively.

The $11/2_1^-$ state in $^{105}$Cd, which has a $T_{1/2}=149$ ps,  
decays via two E1 transitions to two $9/2^+$ states. The partial half-lives
for the two transitions are 2.7$\times 10^{-10}$ and $3.3\times 10^{-10}$ s
respectively.

The $11/2^-$ state in $^{103}$Cd decays via a 931-keV E1 transition to a $9/2^+$. No
time structure of the decaying transition was observed in the present work. 
Therefore, an upper limit of 6 ps was deduced.

The Weisskopf estimates for the 931-keV E1 $\gamma$-ray in $^{103}$Cd is 
$T_{1/2}^{W.e.}=3.8\times 10^{-16}$ s, for the 330-keV E1 transition and 
392-keV E1 $\gamma$-ray in $^{105}$Cd are 8.40$\times 10^{-15}$ s and 
$5.00\times 10^{-15}$ s. and $T_{1/2}^{W.e.}=6.14\times 10^{-12}$ s 
for the 37-keV E1 transition in $^{107}$Cd. For all four transitions 
the E1 hindrance factor $F^{W}=T_{1/2,\gamma}/T_{1/2}^{W.e.}$ is of order of
$10^4$. Given that the 11/2$^-$ is an intruder state, the similar hindrance
factor observed in all three odd-A cadmium nuclei $^{103,105,107}$Cd suggests 
similar structure of the final $9/2^+$ state.

\section{Conclusion}
Excited states in $^{103,105,107}$Cd have been populated via 
fusion-evaporation reactions. Half-lives of several excited states
were measured by using the delayed coincidence technique. 
The half-life of the $7/2_1^+$ state in $^{107}$Cd and $^{105}$Cd were 
confirmed. The half-life of the first excited state in $^{103}$Cd and 
of the $11/2_1^-$ in $^{105}$Cd are newly obtained allowing a systematical study
of the transitions strenghts. 
The B(E2;$7/2_1^+\rightarrow 5/2_1^+$) transitions strenghts in $^{103-107}$Cd 
are strongly hindered with respect to the B(E2;$2^+_1\rightarrow 0^+_1$) values, 
observed in the most deformed nuclei in the region, suggesting a single particle 
nature for the $7/2_1^+$ states. The hindrance factors, calculated for E1 transitions 
in $^{103,105,107}$Cd, suggest similar structure of the $9/2^+$ states.

\section{Acknowledgments}
The work is partly supported by the Bulgarian Science Fund under 
contracts DMU02/1, DRNF02/5, DID-05/16 and by 
a contract for Bularian-Romanian partnership, number BRS-07/23.

\end{document}